\documentclass[aps,prd,eqsecnum,onecolumn,showpacs,nofootinbib,preprintnumbers,a4paper]{revtex4}

    \newcommand*{\be}{\begin{equation}}
    \newcommand*{\ee}{\end{equation}}
    \newcommand*{\ba}{\begin{eqnarray}}
    \newcommand*{\ea}{\end{eqnarray}}
    
    \newcommand*{\LL}[0]{\mathcal{L}}
    \newcommand*{\PP}[0]{\mathcal{P}}

\usepackage[applemac]{inputenc}
\usepackage[T1]{fontenc} 
\usepackage{comment,cancel,amsmath,amsthm,amssymb,braket,graphicx,pifont,hyperref}
\usepackage{natbib}

\begin{document}

\title{Cosmological Perturbations in the Projectable Version of Ho\v{r}ava-Lifshitz
Gravity}

\author{Alessandro \surname{Cerioni}}
\affiliation{Dipartimento di Fisica, Universit\`a degli Studi di
Bologna, via Irnerio 46, I-40126 Bologna, Italy}
\affiliation{INFN, Sezione di Bologna,
via Irnerio 46, I-40126 Bologna, Italy}
\affiliation{Department of Physics, McGill University, Montr\'eal, QC, H3A 2T8, Canada}
\email{cerioni@bo.infn.it, rhb@physics.mcgill.ca}

\author{Robert H.\ \surname{Brandenberger}}
\affiliation{Department of Physics, McGill University, Montr\'eal, QC, H3A 2T8, Canada}
\email{rhb@physics.mcgill.ca}

\begin{abstract}
We consider linear perturbations about a homogeneous
and isotropic cosmological background in the
projectable version of Ho\v{r}ava-Lifshitz gravity. 
Starting from the action for cosmological perturbations, we 
identify the canonically normalized fluctuation variables. We find
that - in contrast to what happens in the non-projectable
version of the theory - the extra scalar cosmological
perturbation mode is already dynamical at the level
of linear perturbations. For values of the parameter
$\lambda$ in the range $1/3 < \lambda < 1$, the extra
mode is ghost-like, for values $1 < \lambda$ and
$\lambda < 1/3$ it is
tachyonic. This indicates a problem for the projectable
version of Ho\v{r}ava-Lifshitz gravity.
\end{abstract}



\date{\today}

\maketitle

\section{Overview}

It has been about one year and a half since Petr Ho\v{r}ava's seminal 
paper first appeared on the arXiv \cite{Horava} (see also \cite{Horava0,Kluson}). 
Ho\v{r}ava proposes  
a model for quantum gravity which is based on 
an anisotropic scaling between space and time, and on constructing the 
most general power-counting renormalizable Lagrangian with respect to this 
new scaling. This construction involves an explicit violation of both general 
covariance and Lorentz  invariance. The construction is modeled after a 
class of condensed-matter models exhibiting anisotropic scaling whose 
prototype is  the theory of a Lifshitz scalar \cite{Lifshitz}.
Hence, the theory is now called ``Ho\v{r}ava-Lifshitz gravity''.

The basic fields in Ho\v{r}ava-Lifshitz (HL) gravity are the same as in Einstein gravity. 
However, there are less symmetries. Hence, it is expected that an extra 
dynamical degree of freedom will emerge. Such a degree of freedom would 
likely cause serious problems for the theory. In particular, it could be 
ghost-like or tachyonic, both of which
would be quite problematic. Indeed, when expanding about
Minkowski space-time, an extra degree of freedom in the scalar gravitational
sector of the theory appears \cite{Horava}. Rather surprisingly, it was found 
\cite{Gao1,Gao2} that in the ``non-projectable'' version of the theory, the extra scalar 
gravitational degree of freedom is non-dynamical when expanding
about a homogeneous and isotropic cosmological background\footnote{There are, however, reasons to expect that the
extra mode will appear beyond linear order in perturbation theory \cite{Blas1}.}.
There are reasons (which will be mentioned in the following section), however,
to prefer the ``projectable'' version of Ho\v{r}ava-Lifshitz gravity.
The structure of the constraints is very different in the two versions of the theory,
and hence it is interesting to study linear cosmological perturbation theory 
in the case of the ``projectable'' version of HL gravity. This is what we do
in this paper, and we 
find that the extra scalar mode of the linear cosmological perturbations becomes
dynamical. The mode is either ghost-like or
tachyonic depending on the value of the parameter $\lambda$ which enters
into the kinetic term of the Ho\v{r}ava-Lifshitz Lagrangian.

The outline of this paper is as follows: in the following section we give an
introduction to the problem and survey previous works. In Section 3 we review
Ho\v{r}ava-Lifshitz gravity and the cosmological background solutions which the
theory admits. Then follows the main section of this paper in which we study
linear cosmological perturbations in the projectable version of the theory,
identify the dynamical variables and study under which conditions the
extra degree of freedom is ghost-like or tachyonic.

\section{Introduction and Review}

Ho\v{r}ava-Lifshitz gravity is based on the assumption that space and time have 
different scaling behaviors with respect to some renormalization group
flow. The flow is defined by a critical exponent $z$:
\ba \label{anisotropic_scaling}
x^i \, &\rightarrow& \, b\, x^i \nonumber \\
t \, &\rightarrow& \, b^z t
\ea
In order to obtain gravity in four dimensional space-time we set $z=3$ so that the 
theory can be power-counting renormalizable, for reasons that 
we will point out later in this paper.
Lorentz invariance is thus explicitly broken if $z\neq 1$, but it is desired that $z$ flow 
to $1$ in the infra-red, thereby restoring Lorentz symmetry at low energies. 
However, it has not yet been shown explicitly that this will occur 
(see \cite{Iengo,Orlando,Wu} for a study of the renormalization group flow 
in scalar field theories of Lifshitz type).

Different versions of HL gravity can be formulated depending on
assumptions other than power-counting renormalizability and
symmetry under both spatial diffeomorphisms and space-independent
time reparametrizations which are added. In particular, one can impose
the requirement of  ``detailed balance'' on the potential terms
in the Lagrangian. Detailed balance was invoked by Ho\v{r}ava in his 
seminal work as a simplifying assumption which reduces the number of
terms and thus simplifies the algebra. However, there
are no particular physical reasons to make this assumption,
and it was also realized that a Lagrangian with detailed
balance leads to phenomenological problems (see e.g.  
\cite{Calcagni,Nastase,Chen1,Charmousis}). 

A more important criterion is ``projectability''. If this condition is imposed, then
the lapse function is independent of space, if not, then it is allowed to depend
on space. Whether or not the projectability condition should be maintained 
is a question which still awaits further investigation. We mention that the 
projectability condition is supported in \cite{Kocharyan} because 
it prevents a non-relativistic theory of gravity from developing 
inconsistencies, whereas, on the other side, it is opposed in 
\cite{Charmousis} because it gives rise to a non-local Hamiltonian constraint, 
and thus appears to make it harder to recover General Relativity (in which
the constraint is local) in the infrared limit\footnote{Note, however,
that a nonlocal constraint can lead to interesting consequences for
cosmology \cite{Shinji2}.}. Another reason in favor of imposing
the projectability condition is that the algebra of constraints appears consistent
only if this condition is imposed \cite{Miao}. The reader is referred to
\cite{Sotiriou} for a more detailed discussion of these points.
More recently, a ``healthy'' extension of HL gravity was proposed
\cite{Blas2}, an extension which may resolve some of the
key challenges the original formulations face.

HL gravity has attracted the attention of cosmologists for several
reasons. It may lead to new solutions of some old cosmological
problems (see e.g. \cite{Kiritsis, Shinji,Calcagni,Chen2}) by providing alternatives to 
the inflationary paradigm of the very early universe. In particular, in a
spatially curved background a bouncing cosmology is obtained
naturally \cite{Brandenberger}, (see also \cite{Leon,Carloni})
and, with the addition of scalar field matter, 
HL gravity can yield a scale-invariant spectrum of curvature perturbations 
in the ultra-violet limit even for non-inflationary expansion of the 
background space-time \cite{Shinji}.

HL gravity starts out with the same number of dynamical degrees
of freedom as General Relativity, but has less symmetries, Specifically,
one loses one gauge symmetry, namely space-dependent
time rescalings. Hence, we should expect an extra degree of freedom to arise
\cite{Horava}. The emergence of this extra degree of freedom has been
studied in a number of works when expanding about flat space-time, i.e.
in the absence of matter (see e.g. 
\cite{Chen1,Cai,Charmousis,Sotiriou,Sfetsos,Bogdanos,Kobakhidze,Park,Kim,Koyama,Maartens1}),
both in the projectable and in the non-projectable versions of the theory.
The extra mode is not only a problem phenomenologically (no such
mode has been observed), but there are also conceptual problems:
the mode is either tachyonic (for values of $\lambda$ in the
ranges  $\lambda < 1/3$ and $1 < \lambda$) or ghost-like (for
$1/3 < \lambda < 1$)\footnote{Note that phenomenological aspects
of the spectrum of gravitational waves \cite{Soda} and
scalar metric fluctuations \cite{Maartens2,Yamaguchi1} have
also been studied.}.

There have been fewer analyses of the extra 
mode in the presence of matter. Interestingly, it was found in \cite{Gao1} 
that in the non-projectable version of the theory, the extra scalar 
gravitational mode is not dynamical
when expanding about a homogeneous and isotropic metric in the
presence of matter\footnote{However, the later work of
\cite{Blas1} indicates that the dangerous extra mode will appear
at higher order in perturbation theory, and also at linear order if the
background is not maximally symmetric.} . 
This conclusion holds both for flat and for
curved spatial sections \cite{Gao2}. In this paper we will show that
in the projectable version of HL gravity the extra mode is
present, and that it has the same tachyonic and ghost-like
properties as a function of $\lambda$ as it does in the absence
of matter when expanding about Minkowski space-time. 

The conclusions we reach are the same as those reached
recently in \cite{Sasaki}. In that paper, the constraint algebra
of linear cosmological perturbation theory about an expanding
cosmological background was analysed, and it was shown why
in the non-projectable version of HL gravity there is only one
dynamical degree of freedom for adiabatic scalar fluctuations,
whereas two remain in the projectable version. Our analysis
reaches the same conclusions, but uses a slightly different
method. It follows that of \cite{Gao1}, in that we work out the 
quadratic action for linear scalar cosmological perturbations, 
identify the dynamical degrees of freedom which have 
canonical normalization, and study the coefficient matrices of 
both the kinetic part of the Lagrangian
and of the mass terms in order to identify possible ghost-like 
or tachyonic states. 

There are other concerns about fluctuations in HL gravity. One is
the issue of strong coupling - which would spoil the recovery of 
General Relativity in the infra-red limit - (see \cite{Charmousis}).
In the infrared limit (which needs to correspond to 
$\lambda \rightarrow 1$ if the theory is to be phenomenologically
viable) the couplings of the extra degree of freedom diverge
in the limit $\lambda \rightarrow 1$ and thus indicate that
cosmological perturbation theory breaks down. This issue is 
controversial even in light of the ``healthy extension'' proposed in \cite{Blas2}
to cure this and other problems of HL gravity (this is further
discussed in  \cite{Papazoglou,Blas3}). Once again, the problem
is absent in linear cosmological perturbation theory in
the case of the non-projectable version of HL gravity \cite{Gao1}.
We will show that the problem reappears in the
projectable version.

In this paper we will consider cosmological perturbations about a 
spatially flat Friedmann-Robertson-Walker (FRW) background in 
the projectable version of HL gravity, without detailed balance, 
and in the presence of scalar field matter. We work out the second 
order action for scalar metric perturbations and study the number of 
physical degrees of freedom and their ghost/tachyonic instabilities. 
The results will be compared to the ones in \cite{Gao1}, where a  
similar study was done in the case of the non-projectable version of
HL gravity (also in the absence of detailed balance), and where it 
was concluded that the extra gravitational degree of freedom which is
dangerous when expanding about flat space-time is non-dynamical,
that the limit $\lambda\rightarrow 1$ - in which General Relativity is 
supposed to be recovered - is smooth, and that no strong-coupling 
problem arises. We find that all of these nice features disappear in
the projectable version\footnote{In work in progress, we are
studying the same questions for the ``safe'' version of HL gravity
of \cite{Blas2}, a version for which some initial work on
cosmological fluctuations has recently been reported in
\cite{Papazoglou,Yamaguchi2}.}.

\section{Setup}

\subsection{Ho\v{r}ava-Lifshitz Gravity}

Ho\v{r}ava-Lifshitz gravity is based on the same metric variables as 
General Relativity. The symmetry group of the theory is assumed to
be different. The theory is not invariant under the full space-time
diffeomorphism group, but only under the restricted group of
spatial diffeomorphisms and space-independent time reparametrizations.
Instead, the anisotropic scaling symmetry of 
(\ref{anisotropic_scaling}) is postulated.

The anisotropy between space and time is readily taken into account 
if we adopt the Arnowitt-Deser-Misner (ADM) decomposition of 
the metric \cite{Arnowitt:1962hi}:
\be\label{ADM_metric}
ds^2 \, = \, -N^2 dt^2 + g_{ij}\left(dx^i + N^i dt\right)\left(dx^j + N^j dt\right) \, ,
\ee
where $t$ is physical time, the $x^i$ coordinates ($i = 1, ..., 3$) are comoving
spatial coordinates, and $g_{ij}$ is the metric on the constant time hypersurfaces.
The gravitational dynamical degrees of freedom are the lapse function 
$N$, the shift vector $N^i$ and the spatial metric $g_{ij}$. In principle 
$N$, $N^i$ and $g_{ij}$ can be functions of both space and time, unless one 
imposes  extra restrictive conditions. In particular, the ``projectability condition'' 
forces $N$ be a function of time only, i.e.  $N \equiv N(t)$.

The construction of the most general action of Ho\v{r}ava-Lifshitz gravity 
relies on the requirement that the theory be power-counting renormalizable
with respect to the scaling symmetry. As reviewed e.g. in \cite{Visser:2009fg},
to obtain the HL version of gravity in $d$ spatial dimensions, we need to set
$z=d$ (see also \cite{Anselmi:2007ri} for a more general analysis). 
Given the choice of $z$, one builds up the action by adding all terms 
which are renormalizable or super-renormalizable, and which are
consistent with the residual symmetries which are imposed. The whole 
procedure is very clearly described in \cite{Sotiriou} and \cite{Chen3} and 
will be only briefly summarized in what follows.

Denoting the scaling dimensions with square brackets and a ``$s$'' subscript, 
we have
\be
[t]_s \, = \, - z, \quad 
[x^k]_s \, =  \, - 1
\ee
as can be seen from Eq.\ (\ref{anisotropic_scaling}). We then require the action 
to be dimensionless. Note that
\be
\left[ S \right]_s \, =  \, \left[\int dt d^d x\, \LL \right]_s \, = \, 0  \,\,\,
\Longleftrightarrow \,\,\, \left[\LL \right]_s \, = \, z+d \, .
\ee
In $3+1$ dimensions the scaling dimension of the Lagrangian must  be 
equal to six, meaning that the Lagrangian is expected to include terms with 
more spatial derivatives terms than those which appear in General Relativity. 
The scaling dimensions of the metric coefficients are
\be
[g_{ij}]_s \, = \, 0, \,\, \quad \,\, [N^i]_s \, = \, z - 1, \,\, \quad \,\, [N]_s \, = \, 0 \, .
\ee
Note that the mass dimensions of both space and time coordinates are still 
equal to $-1$ and hence mass and scaling dimensions do not coincide.

As in \cite{Maartens1,Maartens2}, we consider the following action:
\be
S \, = \, \chi^2 \int dt d^3 x N \sqrt{g} \left( \LL_K - \LL_V + \chi^{-2} \LL_M \right)
\ee
where $g \equiv  \det(g_{ij})$, $\chi^2 \equiv 1/(16\pi G)$ and the kinetic, 
potential and matter Lagrangians are, respectively, given by
\begin{subequations}
\begin{align}
\LL_K &= K_{ij}K^{ij}-\lambda K^2\label{kinetic}\\
\LL_V &= 2\Lambda-R+\frac{1}{\chi^2}(g_2 R^2 +g_3 R_{ij}R^{ij}) + \frac{1}{\chi^4}\left(g_4 R^3 + g_5 R R_{ij} R^{ij} + g_6 R^i_j R^j _k R^k_i \right)+ \\
& +\frac{1}{\chi^4}\left[ g_7 R \nabla^2 R + g_8 (\nabla_i R_{jk})(\nabla^i R^{jk})\right]\nonumber\\
\LL_M &= \frac{1}{2N^2}\left(\dot\varphi-N^i\nabla_i \varphi \right)^2-V(g_{ij},\mathcal{P}_n,\varphi) \, ,
\end{align}
\end{subequations}
and the extrinsic curvature $K_{ij}$ is defined as
\be
K_{ij} = \frac{1}{2N}\left(-\dot g_{ij}+\nabla_i N_j + \nabla_j N_i\right) \, ,
\ee
and, as customary, dots over variables represent time derivatives. Indices are 
raised and lowered using the spatial metric $g_{ij}$ and all the terms in the  
action depending on the Ricci tensor and its contractions are built from the 
same spatial metric. The scalar field Lagrangian differs from that in
theories with full general covariance in that it can also contain terms
with higher spatial derivatives as long as they do not destroy the
power-counting renormalizability of the theory with respect to the scaling
symmetry. Thus, the non-kinetic part of the scalar field
Lagrangian can be written as follows \cite{Maartens1,Maartens2,Chen3}:
\be
V = V_0(\varphi) + V_1(\varphi) \mathcal{P}_0 + V_2(\varphi) \mathcal{P}_1^2 + V_3(\varphi) \PP_1^3 + V_4(\varphi) \PP_2 +V_5(\varphi) \PP_0\PP_2 + V_6(\varphi)\PP_1\PP_2 \, ,
\ee
where
\be
\mathcal{P}_0 \equiv (\nabla\varphi)^2, \, \mathcal{P}_i \equiv \Delta^i \varphi, \, \Delta \equiv g^{ij} \nabla_i \nabla_j \, .
\ee

Some remarks are in order:
\begin{enumerate} 
\item
If one introduces the ADM metric (\ref{ADM_metric}) into the Einstein-Hilbert action, 
\be 
S_{EH} = \chi^2 \int d^4 x\, \sqrt{-g}\, {^{(4)}R}
\ee
where $^{(4)}R$ is the four-dimensional Ricci scalar, the resulting kinetic 
Lagrangian will have the same form as (\ref{kinetic}) with $\lambda=1$, 
whereas the potential Lagrangian would simply include the term 
$\propto R$ (see for instance  \cite{DeWitt:1967yk}).

\item 
Note that there are several coupling constants entering the potential contribution
$\LL_V$ to the Lagrangian, namely $\chi$ and the $g$'s.  The coupling constants 
$g_I$, $I=(2,\dots,8)$, have zero mass dimension, but non-zero scaling dimensions, 
while $\chi$ is dimensionless with respect to the scaling symmetry but has 
energy dimension equal to one. We further observe that the couplings to 
$\Lambda$ and $R$ are set to unity. These two terms
are super-renormalizable with respect to the scaling symmetry.
\end{enumerate}

Varying the action with respect to $N(t)$ one can derive the following 
Hamiltonian constraint:
\be\label{hamiltonian_constraint}
\int d^3 x \sqrt{g} \left(\LL_K  + \LL_V - \frac{1}{2\chi^2}J^t \right)= 0 \, ,
\ee
where
\be
J^t \equiv 2 \left(N \frac{\delta\LL_M}{\delta N} + \LL_M \right) \, .
\ee
The Hamiltonian constraint is non-local in the projectable version
of HL gravity (it is expressible as a volume integral 
instead of being local). 

The variation of the action with respect to $N^i(t,x^k)$ results in the 
following super-momentum constraint:
\be\label{super-momentum_constraint}
\nabla_i \pi^{ij} = \frac{1}{2\chi^2} J^j \, ,
\ee
where the super-momentum $\pi^{ij}$ and the matter current are given, 
respectively, by:
\be
\pi^{ij} \equiv \frac{\delta\LL_K}{\delta \dot g_{ij}} = -K{^{ij}} + \lambda K g_{ij}
\ee
and
\be
J_i \equiv - N \frac{\delta \LL_M}{\delta N^i} = \frac{1}{N}\left(\dot\varphi-N^k \nabla_k \varphi\right)\nabla_i \varphi \, .
\ee
We will make explicit both constraints in the following section.

\subsection{Cosmological Background}

The metric in the ADM form as in Eq.\ (\ref{ADM_metric}) can be reduced to
a spatially flat FRW metric if one sets the values of $N$, $N^i$ 
and $g_{ij}$ as follows:
\ba
N(t) \, &=& \, 1+\delta N(t),\nonumber \\ 
N^i(t,x^k) \, &=& \, 0 +\delta N^i(t,x^k), \\ 
g_{ij}(t,x^k) \, &=& \,  a^2(t)\delta_{ij} + \delta g_{ij}(t,x^k) \, . \nonumber
\ea
The perturbations $\delta N$, $\delta N^i$, $\delta g_{ij}$ will be specified later on.

On evaluating the Hamiltonian constraint Eq.\ (\ref{hamiltonian_constraint}) 
to zeroth order one obtains
\be\label{Friedmann}
(3\lambda-1)\,H^2 \, = \, \frac{1}{3}\left(\frac{\rho_M}{\chi^2} + 2\Lambda  \right)
\ee
which generalizes the first Friedmann equation. Here, $\rho_M$ is the energy 
density associated to the matter sector. Note that in the absence of $\Lambda$, 
$H^2$ is strictly positive only for $\lambda > 1/3$. Otherwise, if  
$\lambda<1/3$, it is positive only in presence of a sufficiently negative 
cosmological constant ($\Lambda < -\rho_M / M_\textrm{Pl}$). However, 
the range $\lambda<1/3$ is not interesting phenomenologically  
because it is disconnected by the singular point
$\lambda = 1/3$ from the value $\lambda=1$ for which one wishes to 
recover General Relativity.

The dynamical equation for the scale factor $a(t)$, the generalization of 
the second Friedmann equation, can be obtained by varying the action 
with respect to $g_{ij}$ and evaluating the result in the homogeneous limit. 
The result is:
\be\label{dyn_for_a}
(3\lambda-1)\frac{\ddot a}{a} = -\frac{1}{6\chi^2} (\rho_M + 3p_M) + \frac{2}{3}\Lambda \, ,
\ee
where $p_M$ is the (background) pressure associated with matter. 
For scalar field matter we have
\ba \label{SF_energy_density_and_pressure}
\rho_M \, &=& \,  \frac{\dot\varphi_0^2}{2} + V_0(\varphi_0), \nonumber \\
p_M \, &=& \,  \frac{\dot\varphi_0^2}{2} - V_0(\varphi_0) \, ,
\ea
and the background equation of motion becomes:
\be \label{eom}
\ddot\varphi_0+3H\dot\varphi_0 \, = \,  -\frac{dV_0(\varphi_0)}{d\varphi_0}
\ee
exactly as in General Relativity.

\section{Cosmological Perturbations}

\subsection{Introduction}

In this section we are interested in scalar metric fluctuations, the
fluctuation modes which couple to energy density and pressure.
Vector and tensor perturbations are studied in \cite{Sasaki}.
The basic fluctuation variables are the same as in the case
of General Relativity (see e.g. \cite{MFB} for an in-depth
review of the theory of cosmological fluctuations and
\cite{RHBrev} for an introductory overview):
\begin{subequations} \label{metric_pert}
\begin{align}
\delta N(t) &= \phi(t)\\
\delta N_i(t,x^k) &= \partial_i B(t,x^k)\\
\delta g_{ij}(t,x^k) &= a^2(t) \left[-2\,\psi(t,x^k)\,\delta_{ij} + 2\,E(t,x^k)_{|ij}\right] 
\end{align}
\end{subequations}
where the subscript ${}_{|i}$ denotes the covariant derivative.
Correspondingly, also matter fluctuations must be taken into account:
\be \label{matter_pert}
\varphi(t,x^k) = \varphi_0(t) + \delta\varphi(t,x^k)\, .
\ee

The first major difference (particular to the projectable version of HL gravity)
compared to Einstein gravity is that the variable $\phi$ depends only
on time. The second major difference concerns the symmetry group.
In the case of HL gravity it is reduced compared to the case of Einstein
gravity. One loses the space-dependent time reparametrizations.
In the case of General Relativity one can use the two scalar gauge
degrees of freedom to set $E = B = 0$. In HL gravity there is only
one space-dependent gauge mode. One can use this mode to
realize the gauge choice $E = 0$. In the projectable version
of HL gravity one can in addition make use of space-independent
time reparametrizations to set $\phi = 0$ (in the non-projectable
version of the theory one cannot make this choice since $\phi$ then
can depend non-trivially on space). In conclusion, in our work
can use the gauge freedom to set
\be
\phi \, = \, 0, \,\,\, \quad \,\,\, E \, = \, 0 \, ,
\ee
which corresponds to the choice of the quasi-longitudinal gauge (see
also \cite{Maartens1,Maartens2}).

Expanding the Hamiltonian constraint (\ref{hamiltonian_constraint}) 
to first order one finds
\be
\int d^3 x \; a^3 \left[ 2\Delta \psi - (3\lambda-1) H (\Delta B + 3\dot\psi) -\frac{\delta\rho_M}{2\chi^2} \right] = 0 \, ,
\ee
which is actually trivially satisfied when dealing with linear cosmological 
perturbations since the spatial average of linear fluctuations must vanish 
(any non-vanishing term would be a contribution to the background solution).
Note at this stage a key difference compared to the non-projectable version
of HL gravity: in the latter case the linear Hamiltonian constraint is
local and hence provides non-trivial constraints.

In contrast, the super-momentum constraint  (\ref{super-momentum_constraint}) 
is trivial at the homogeneous level but not at first order where is becomes
\be \label{1st_SM_cnstr}
\partial_j\left[(\lambda-1)\Delta B + (3\lambda-1) \dot\psi - \frac{1}{2\chi^2}q_M \right] = 0 \, ,
\ee
with
\be
q_M = \dot\varphi_0\delta\varphi \, .
\ee
In linear perturbation theory we can work in Fourier space where the spatial 
derivative $\partial_i$ can be replaced by $(-i \,k_j)$. Hence, the quantities 
inside the square brackets of (\ref{1st_SM_cnstr}) must sum to zero.

Note that we started with five scalar degrees of freedom as in 
Eqs.\ (\ref{metric_pert}), (\ref{matter_pert}), and then we have 
decreased their number by two by making use of the gauge freedom. 
The number of degrees of freedom can be further reduced by one using the 
first order super-momentum constraint (\ref{1st_SM_cnstr}) to remove $B$ 
and thus only two  degrees of freedom - $\psi$ and $\delta\varphi$ - are left.
 
\subsection{Second-order action}

In the following we will insert the ansatz for cosmological fluctuations
of the previous subsection into the action for HL gravity and 
determine the second order action, the terms quadratic in the
fluctuation variables. This will allow us to find the canonically
normalized fluctuation variables and determine if they
are ghost-like or tachyonic. 

The second order action receives three contributions, namely:
\be
\delta_2 S^{(s)} = \chi^2 \int dt d^3 x\, \left[ \delta_0(\sqrt{g})\, \delta_2\LL ^{(s)}+ \delta_1 (\sqrt{g}) \delta_1\LL^{(s)} + \delta_2 (\sqrt{g})\delta_0\LL^{(s)}\right] \equiv   \chi^2 \int dt d^3 x\,  a^3 \,\LL_2 ^{(s)} \, ,
\ee
where we have implicitly introduced the following notation to denote the orders in the perturbative expansion:
\be
f \equiv \sum_{i=0}^{\infty}\delta_i f \, .
\ee
Concerning the expansion of $\sqrt{g}$ one readily has
\be
\delta_0(\sqrt{g}) = a^3, \quad \delta_1(\sqrt{g}) 
=-3 a^3 \psi,\quad \delta_2 (\sqrt{g}) = \frac{3}{2} a^3 \psi^2 \, .
\ee

After making use of the gauge choice to eliminate $\phi$ and $E$, and
the constraint equation to express $B$ in terms of the two remaining
scalar degrees of freedom, the second order scalar action acquires the 
following form in terms of $\psi$ and $\delta\varphi$:
\be\label{2nd_order_action_psi_deltaphi}
\begin{split}
\LL_2^{(s)}[\psi,\delta\varphi] = & \frac{4(3\lambda-1)}{(\lambda-1)}\frac{\dot\psi^2}{2} + \frac{\dot{\delta\varphi}^2}{2\chi^2}  + f_\psi \psi \dot\psi +f_{\varphi\psi}\psi\dot{\delta\varphi}+\tilde f_{\varphi\psi}\dot\psi\delta\varphi+ \\
&- m_\psi^2 \psi^2 - m_\varphi^2 \delta\varphi^2 - m_{\varphi\psi}^2\psi\delta\varphi\\
& + \omega_\varphi \delta\varphi \Delta\delta\varphi + \omega_\psi \psi \Delta\psi \\
&+  d_\psi (\Delta\psi)^2+d_\varphi (\Delta\delta\varphi)^2 + \tilde d_\psi \Delta \psi \Delta^2 \psi+\tilde d_\varphi \Delta \delta\varphi \Delta^2 \delta\varphi
\end{split}
\ee
where the various coefficients are listed in Appendix \ref{coeff_psi_deltaphi}. In 
order to obtain this result we did some integrations by parts in intermediate 
steps and used the background dynamical equations for $a(t)$ and $\varphi_0(t)$.

We observe that the coefficient multiplying $\dot\psi^2$ has a ``wrong'' 
negative sign for $1/3 < \lambda < 1$, which will give rise to ghost 
instabilities as reported in almost all the literature about Ho\v{r}ava-Lifshitz 
gravity (see for instance \cite{Bogdanos,Maartens1,Koyama,Sotiriou,Chen2}).

In order to compare our result with the analyses in previous works done
in the absence of matter, we can set the matter terms to zero in our
result and consider the remaining pieces in the second order action 
for the extra gravitational scalar degree of freedom which reads as follows:
\be
\begin{split}
\delta_2 S^{(s)}[\psi] =& \int dt d^3 x\,a^3\,\Bigg\{\frac{4(3\lambda-1)}{(\lambda-1)}\frac{\dot\psi^2}{2}  + 6H(1-3\lambda) \psi \dot\psi - 15 (1-3\lambda)H^2 \psi^2 + \\
& -2\psi\Delta\psi -(16g_2+6g_3)\frac{(\Delta\psi)^2}{\chi^2} + (6g_8-16g_7)\frac{\Delta\psi\Delta^2\psi}{\chi^4}\Bigg\} \, .
\end{split}
\ee
This  result is very similar to Eq.\ (33) of Ref.\ \cite{Chen2} and to 
Eq.\ (39) of Ref.\ \cite{Chen1} once the spatial derivatives are set to zero, 
except for a discrepancy in the coefficient multiplying 
$(1-3\lambda)H^2 \psi^2$ which is $-15$ in our result instead of
$27$ appearing in the cited references.

In order to draw definite conclusions about the ghost nature of the
fluctuation modes, we must identify the canonically normalized
variables. For values of $\lambda$ which lie in the regions
$\lambda < 1/3$ or $\lambda > 1$ we rescale the fields as
\be
{\tilde\psi}\equiv \sqrt{\frac{4(3\lambda-1)}{\lambda-1}}\psi,\quad \tilde{\delta\varphi}\equiv \frac{\delta\varphi}{\chi}
\ee
and obtain the following second order Lagrangian in terms of canonically 
normalized variables:
\be\label{2nd_order_action_tilde_psi_tilde_deltaphi}
\begin{split}
\LL_2^{(s)}[\psi,\delta\varphi] = & \frac{1}{2}{\dot{\widetilde\psi}}^2 + \frac{1}{2}\dot{\widetilde{\delta\varphi}}^2  + f_{\widetilde\psi} \widetilde\psi \dot{\widetilde\psi} +f_{\widetilde\varphi\widetilde\psi}\widetilde\psi\dot{\widetilde{\delta\varphi}}+ \tilde f_{\widetilde\varphi\widetilde\psi}\dot{\widetilde\psi}\widetilde{\delta\varphi}+ \\
&- m_{\widetilde\psi}^2 \widetilde\psi^2 - m_{\widetilde\varphi}^2 \widetilde{\delta\varphi}^2 - m_{\widetilde\varphi\tilde\psi}^2\widetilde\psi\widetilde{\delta\varphi}\\
& + \omega_{\widetilde\varphi} \widetilde{\delta\varphi} \Delta\widetilde{\delta\varphi} + \omega_{\widetilde\psi} \widetilde\psi \Delta \widetilde\psi \\
&+  d_{\widetilde\psi} (\Delta \widetilde\psi)^2+d_{\widetilde\varphi} (\Delta \widetilde{\delta\varphi})^2  +  \tilde d_ {\widetilde\psi} \Delta \widetilde \psi \Delta^2 \widetilde\psi+\tilde d_{\widetilde\varphi} \Delta \widetilde{\delta\varphi} \Delta^2 \widetilde{\delta\varphi}\, .\end{split}
\ee
The coefficients can be found in Appendix \ref{coeff_tilde_psi_tilde_deltaphi}.
Thus, the degrees of freedom have a positive sign for the kinetic terms in the action
and there are no ghosts. In the range $1/3 < \lambda < 1$ we must use
the rescalings
\be
{\tilde\psi}\equiv \sqrt{\frac{-4(3\lambda-1)}{\lambda-1}}\psi,\quad \tilde{\delta\varphi}\equiv \frac{\delta\varphi}{\chi}
\ee
which changes the sign of the kinetic term of $\psi$. Hence, in this range of
values of $\lambda$ the extra gravitational degree of freedom is
ghost-like.

\subsection{Number of physical degrees of freedom}

In  \cite{Gao1} perturbations in the non-projectable version of Ho\v{r}ava-Lifshitz 
gravity were analyzed, and it was shown that not all the degrees of freedom 
which na\"ively appear in an expansion similar to 
Eq.\ (\ref{2nd_order_action_psi_deltaphi}) are really dynamical. 
Indeed, after introducing the Sasaki-Mukhanov \cite{Sasaki2,Mukh} 
variable $\zeta$ defined as
\be
\zeta \equiv - \psi - \frac{H}{\dot\varphi_0}\delta\varphi
\ee
and substituting for $\delta\varphi$ in terms of $\zeta$, there 
remained only one variable and it entered the Lagrangian 
with a proper kinetic term. Thus, the potentially dangerous
degree of freedom was in fact not dynamical. The same 
``trick'' is not successful in the present case. Written in
terms of $\zeta$, the Lagrangian takes the following form:
\be
\begin{split}\label{2nd_order_action_psi_zeta}
\LL_2^{(s)}[\psi,\zeta] =&  \frac{\dot\varphi_0^2}{H^2 \chi^2}\frac{\dot\zeta^2}{2} + \left[\frac{4(3\lambda-1)}{(\lambda-1)}+\frac{\dot\varphi_0^2}{H^2\chi^2} \right]\frac{\dot\psi^2}{2} +f_\zeta \zeta \dot\zeta+ f_\psi \psi \dot\psi +f_{\zeta\psi}\psi\dot\zeta+\tilde f_{\zeta\psi}\zeta\dot\psi+ g_{\zeta\psi}\dot\zeta\dot\psi +\\
&- m_\psi^2 \psi^2 - m_\zeta^2 \zeta^2 - m_{\zeta\psi}^2\zeta\psi\\
& + \omega_\zeta \zeta \Delta\zeta + \omega_\psi \psi \Delta\psi + \omega_{\zeta\psi} \psi \Delta\zeta + \tilde \omega_{\zeta\psi}\zeta\Delta\psi +\\
&+  d_\psi (\Delta\psi)^2+d_\zeta (\Delta\zeta)^2 + d_{\zeta\psi}\Delta\zeta\Delta\psi+\tilde d_{\zeta\psi} \Delta\zeta \Delta^2 \psi + \tilde d_\psi \Delta \psi \Delta^2 \psi+\tilde d_\zeta \Delta \zeta \Delta^2 \zeta \, .
\end{split}
\ee
Once again, the various coefficients are listed in Appendix \ref{coeff_psi_zeta}. 
Observe that, even in absence of any matter field, $\psi$ is still a 
dynamical (gravitational) degree of freedom. 

We wish to emphasize the fact that - as opposed to the situation in the
non-projectable version - in the projectable version of HL gravity 
it is not possible to reduce to one the number of physical degrees of freedom, 
in agreement with the results of the general analysis of \cite{Sasaki} and with 
the conclusions reached in many other 
papers in which perturbations around Minkowski background are analyzed. 
Note that the difference in the number of physical degrees of freedom of
linear cosmological perturbations between the non-projectable version
of HL gravity (analyzed in  \cite{Gao1}) and the projectable version
analyzed here is in complete agreement with the analysis made in  
\cite{Sasaki} based on the classification of constraints in the Hamiltonian formalism.

\subsection{Mass eigenvalues and discussion on tachyonic instabilities}

We now want to investigate the issue of tachyonic (classical) instabilities.
We do this by looking at the signs of the eigenvalues of the mass matrix.
In general it is difficult to diagonalize the mass matrix. Hence, 
we will specialize to a couple of cases, both of them with $\Lambda = 0$  
and for a potential $V_0(\varphi_0) = m^2 \varphi_0^2/2$. The first example will
be a static field, the second a field oscillating around $\varphi_0=0$.

\subsubsection{Static field and $\Lambda = 0$}

Setting $\varphi_0 = x \chi$, where $x$ is a dimensionless constant,
the mass terms in 
Appendix \ref{coeff_tilde_psi_tilde_deltaphi} read as follows:
\begin{subequations}
\begin{align}
m_{\widetilde\psi}^2  &= -\frac{5}{8}\frac{\lambda-1}{3\lambda-1}m^2 x^2\\
m_{\widetilde{\varphi}}^2  &= \frac{m^2}{2}\\
m_{\widetilde{\varphi}\widetilde\psi}^2 &= -\frac{3}{2}\sqrt{\frac{\lambda-1}{3\lambda-1}}m^2 x \, .
\end{align}
\end{subequations}
The mass matrix defined as
\be
\widetilde M^2_{\widetilde\varphi\widetilde\psi} \equiv \left(\begin{array}{c c} m_{\widetilde\varphi}^2 & m_{\widetilde\varphi\widetilde\psi}^2/2 \\  m_{\widetilde\varphi\widetilde\psi}^2/2 & m_{\widetilde\psi}^2 \end{array} \right)
\ee
can be easily diagonalized and its eigenvalues are
\begin{subequations}
\begin{align}
\widetilde M_+^2 &= \frac{m^2}{16(3\lambda-1)}\left[4(3\lambda-1) - 5x^2(\lambda-1)+\sqrt{16(3\lambda-1)^2 + 25x^4 (\lambda-1)^2 + 184 x^2 (\lambda-1)(3\lambda-1)} \right]
\\
\widetilde M_-^2 &= \frac{m^2}{16(3\lambda-1)}\left[4(3\lambda-1) - 5x^2(\lambda-1)-\sqrt{16(3\lambda-1)^2 + 25x^4 (\lambda-1)^2 + 184 x^2 (\lambda-1)(3\lambda-1)} \right] \, .
\end{align}
\end{subequations}
In both of the ranges $\lambda > 1$ and $\lambda < 1/3$ one eigenvalue is 
positive ($\widetilde M_+^2$) whereas the other one is negative 
for any value of the scalar field. Thus, the extra scalar metric degree of
freedom has a tachyonic instability in these regions of $\lambda$ (the ones
which do not suffer from the ghost problem), as is also known from 
previous works which considered fluctuations in a theory without
matter.

In terms of the variables $\zeta$, $\psi$ we obtain the following eigenvalues:
\begin{subequations}
\begin{align}
M_+^2 &= -\frac{m^2}{4}\left[12(3\lambda-1) + 5 x^2 + \sqrt{25x^4 + 144 (3\lambda-1)^2} \right]\\
M_-^2 &=- \frac{m^2}{4}\left[12(3\lambda-1) + 5 x^2 - \sqrt{25x^4 + 144 (3\lambda-1)^2} \right]
\end{align}
\end{subequations}
which are both negative for any $\lambda>1/3$ and for any $x$.

\subsubsection{Oscillating field and $\Lambda = 0$}

We set $\varphi_0 = A \cos (m t)$ and then average over field oscillations as follows:
\be
\langle f(t) \rangle \equiv \frac{m}{2\pi }\int _{-\pi/m}^{\pi/m} dt \, f(t) \, .
\ee
For instance, we get the following result for the Hubble parameter,
\be
\langle H^2 \rangle = \frac{1}{3\chi^2 (3\lambda-1)}\frac{m^2 A^2}{2} \, ,
\ee
while the mass terms become
\begin{subequations}
\begin{align}
\langle m_{\widetilde\psi}^2 \rangle &= -\frac{13}{6}\frac{\lambda-1}{3\lambda-1}\frac{m^2 A^2}{\chi^2}\\
\langle m_{\widetilde{\varphi}}^2 \rangle &= \frac{m^2}{8}\left(4-\frac{1}{\lambda-1}\frac{A^2}{\chi^2} \right)\\
\langle m_{\widetilde\psi\widetilde{\varphi}}^2 \rangle  &= 0 \, .
\end{align}
\end{subequations}
We see that $\langle m_{\widetilde\psi}^2 \rangle$ is negative in both of the
regions  $\lambda > 1$ and $\lambda < 1/3$ which are ghost-free.
Thus, $\widetilde\psi$ displays tachyonic instability.

In terms of $\zeta$ and $\psi$ we get
\begin{subequations}
\begin{align}
\langle m_\zeta^2 \rangle &= -\frac{9m^2 A^2 (7\lambda-3)}{32(\lambda-1)\chi^2}\\
\langle m_\psi^2 \rangle &= -\frac{m^2 A^2 (131\lambda-95)}{32(\lambda-1)\chi^2}\\
\langle m_{\zeta\psi}^2 \rangle &= -\frac{9m^2 A^2 (5\lambda-1)}{16(\lambda-1)\chi^2}
\end{align}
\end{subequations}
and the following eigenvalues, which are both negative for any $\lambda>1$:
\be
M_+^2 = -\frac{m^2 A^2}{32(\lambda-1)\chi^2}\left(97\lambda-61+\sqrt{1237-3122\lambda+3181\lambda^2} \right) \, ,
\ee
\be
M_-^2 = -\frac{m^2 A^2}{32(\lambda-1)\chi^2}\left(97\lambda-61-\sqrt{1237-3122\lambda+3181\lambda^2} \right) \, .
\ee

\section{Conclusions}

In this paper we have studied linear cosmological perturbations in the projectable
version of Ho\v{r}ava-Lifshitz gravity. We find that - unlike what happens in the
non-projectable version - there are two physical degrees of freedom for scalar
metric fluctuations (in addition to possible entropy modes). The extra degree
of freedom is either ghost-like (for $1/3 < \lambda < 1$) or tachyonic 
(for $\lambda < 1/3$ and $\lambda > 1$). Hence, linear cosmological
perturbation theory is sick for all values of $\lambda$ except for the
value $\lambda = 1$ corresponding to Einstein gravity.

Turning to the ``strong coupling problem'' first discussed in \cite{Charmousis},
we notice that - again in contrast to what happens in the non-projectable
version of HL gravity - the strong coupling problem in the limit 
$\lambda \rightarrow 1$ manifests itself in the divergence of the coefficients
in the second order action for the extra degree of freedom. Tracing back
the origin of this divergence, we see that it comes from the factor
$(\lambda - 1)$ which multiplies the variable $B$ in the super-momentum
constraint. The super-momentum constraint is used to solve for $B$, and
hence a divergence arises in the limit $\lambda \rightarrow 1$. In
General Relativity, $B$ is a pure gauge mode and can be set to zero from
the outset. In the non-projectable version of HL gravity a combination
of the super-momentum and Hamiltonian constraints is used to solve
for $B$, and in this combination the limit $\lambda \rightarrow 1$ does
not lead to any strong coupling instability if $H \neq 0$.

In conclusion, we find that cosmological perturbation theory in the
projectable version of HL gravity suffers from the presence of an
extra unwanted degree of freedom for scalar metric fluctuations.
To make matters worse, this degree of freedom is either ghost-like
of else has a tachyonic instability (depending on the value of
$\lambda$. Furthermore, in the limit $\lambda \rightarrow 1$
a strong coupling instability arises. In work in progress we are
studying linear cosmological perturbations in the ``safe''
version of HL gravity proposed in \cite{Blas2}, with the goal
of checking if any of the problems discussed above are cured.

\begin{acknowledgments}

This work is supported in part by a NSERC Discovery Grant, by funds from
the CRC Program and by a Killam Research Fellowship awarded to R.B. 
We wish to thank Shinji Mukoyhama, Yi Wang and Wei Xue for useful discussions.

\end{acknowledgments}

\appendix
\section{Coefficients in Eq.\ (\ref{2nd_order_action_psi_deltaphi})}\label{coeff_psi_deltaphi}
\be
f_\psi = -6H(3\lambda-1),\quad f_{\varphi\psi} =-3\frac{\dot\varphi_0}{\chi^2},\quad \tilde{f}_{\varphi\psi} = -\frac{3\lambda-1}{\lambda-1}\frac{\dot\varphi_0}{\chi^2}
\ee
\begin{subequations}
\begin{align}
m_\psi^2 &= -\frac{39}{2}(3\lambda-1)H^2  + 3\Lambda + \frac{3}{2}\frac{V_0(\varphi_0)}{\chi^2}-\frac{3}{4}\frac{\dot\varphi_0^2}{\chi^2} \\
m_\varphi^2 &= -\frac{1}{4(\lambda-1)}\frac{\dot\varphi_0^2}{\chi^4}+\frac{1}{2}\frac{V_{0,\varphi\varphi}(\varphi_0)}{\chi^2}\\
m_{\psi\varphi}^2 &= -3\frac{V_{0,\varphi}(\varphi_0)}{\chi^2}
\end{align}
\end{subequations}
\be
w_\psi = -2,\quad w_\varphi= \frac{V_1(\varphi_0)}{\chi^2}
\ee

\be
d_\varphi = - \frac{V_{4,\varphi}(\varphi_0)}{\chi^2}-\frac{V_2(\varphi_0)}{\chi^2}, \quad d_\psi = -16\frac{g_2}{\chi^2}-6\frac{g_3}{\chi^2},\quad \tilde d_\psi = 6\frac{g_8}{\chi^4}-16\frac{g_7}{\chi^4},\quad \tilde d_{\widetilde\varphi}= -\frac{V_6(\varphi_0)}{\chi^2}
\ee

\section{Coefficients in Eq.\ (\ref{2nd_order_action_tilde_psi_tilde_deltaphi})}\label{coeff_tilde_psi_tilde_deltaphi}
\be
f_{\widetilde\psi} =-\frac{3}{2}H(\lambda-1) ,\quad f_{\widetilde\varphi\widetilde\psi} =-\frac{3}{2}\sqrt{\frac{\lambda-1}{3\lambda-1}}\frac{\dot\varphi_0}{\chi},\quad \tilde{f}_{\widetilde\varphi\widetilde\psi} = -\frac{1}{2}\sqrt{\frac{3\lambda-1}{\lambda-1}}\frac{\dot\varphi_0}{\chi}
\ee

\begin{subequations}
\begin{align}
m_{\tilde\psi}^2  &= -\frac{3}{16}\frac{\lambda-1}{3\lambda-1}\frac{\dot\varphi_0^2}{\chi^2} +\frac{3}{8}\frac{\lambda-1}{3\lambda-1}\frac{V_0(\varphi_0)}{\chi^2}- \frac{39}{8}(\lambda-1)H^2 +\frac{3}{4}\frac{\lambda-1}{3\lambda-1}\Lambda\\
m_{\tilde{\delta\varphi}}^2 &= -\frac{1}{4(\lambda-1)}\frac{\dot\varphi_0^2}{\chi^2}+\frac{1}{2}V_{0,\varphi\varphi}(\varphi_0)\\
m_{\tilde\psi\tilde{\delta\varphi}}^2  &= -\frac{3}{2}\sqrt{\frac{\lambda-1}{3\lambda-1}}\frac{V_{0,\varphi}(\varphi_0)}{\chi}
\end{align}
\end{subequations}

\be
w_{\widetilde\psi} = -\frac{1}{2}\frac{\lambda-1}{3\lambda-1}, \quad \omega_{\widetilde\varphi} = V_1(\varphi_0)
\ee

\be
d_{\widetilde\varphi} = -V_{4,\varphi}(\varphi_0) -V_2(\varphi_0), \quad  d_{\widetilde\psi} = -4\frac{\lambda-1}{3\lambda-1}\frac{g_2}{\chi^2}-\frac{3}{2}\frac{\lambda-1}{3\lambda-1}\frac{g_3}{\chi^2},\quad \tilde d_{\widetilde\psi} = -4\frac{\lambda-1}{3\lambda-1}\frac{g_7}{\chi^4}+\frac{3}{2}\frac{\lambda-1}{3\lambda-1}\frac{g_8}{\chi^4},\quad \tilde d_{\widetilde\varphi} = -V_6(\varphi_0)
\ee

\section{Coefficients in Eq.\ (\ref{2nd_order_action_psi_zeta})}\label{coeff_psi_zeta}
In what follows it is
\be
F(\varphi_0) \equiv 2\Lambda \chi^2 - \dot\varphi_0^2 + V_0(\varphi_0)
\ee

\begin{subequations}
\begin{align}
m_\zeta^2  = &  - \frac{1}{4(\lambda-1)}\frac{\dot\varphi_0^4}{\chi^4 H^2} - \frac{1}{2}\frac{\dot\varphi_0^2}{\chi^2}+\frac{1}{2}\frac{V_{0,\varphi\varphi}\dot\varphi_0^2}{\chi^2 H^2}+\nonumber\\
&-\frac{1}{3(3\lambda-1)}\frac{\dot\varphi_0}{\chi^4 H^3}(3H\dot\varphi_0+V_{0,\varphi})F(\varphi_0)-\frac{1}{2}\frac{(3H\dot\varphi_0+V_{0,\varphi})^2}{\chi^2 H^2}+\frac{\dot\varphi_0}{\chi^2 H}(3H\dot\varphi_0+V_{0,\varphi})\nonumber\\
&-\frac{1}{18(3\lambda-1)^2}\frac{\dot\varphi_0^2}{\chi^6 H^4}F(\varphi_0)^2+\frac{1}{3(3\lambda-1)}\frac{\dot\varphi_0^2}{\chi^4 H^2}F(\varphi_0)\\
m_\psi^2 =& -\frac{39}{2} H^2 (3\lambda-1)+3\Lambda +3\frac{\dot\varphi_0}{H\chi^2}V_{0,\varphi}- \frac{1}{4(\lambda-1)}\frac{\dot\varphi_0^4}{\chi^4 H^2} +\frac{1}{2}\frac{V_{0,\varphi\varphi}\dot\varphi_0^2}{\chi^2 H^2}-\frac{17}{4}\frac{\dot\varphi_0^2}{\chi^2}+\frac{3}{2}\frac{V_0(\varphi_0)}{\chi^2}\nonumber\\
&-\frac{1}{3(3\lambda-1)}\frac{\dot\varphi_0}{\chi^4 H^3}(3H\dot\varphi_0+V_{0,\varphi})F(\varphi_0)-\frac{1}{2}\frac{(3H\dot\varphi_0+V_{0,\varphi})^2}{\chi^2 H^2}+4\frac{\dot\varphi_0}{\chi^2 H}(3H\dot\varphi_0+V_{0,\varphi})\nonumber\\
&-\frac{1}{18(3\lambda-1)^2}\frac{\dot\varphi_0^2}{\chi^6 H^4}F(\varphi_0)^2+\frac{4}{3(3\lambda-1)}\frac{\dot\varphi_0^2}{\chi^4 H^2}F(\varphi)\\
m_{\zeta\psi}^2 =& +\frac{5}{3(3\lambda-1)}\frac{\dot\varphi_0^2}{\chi^4 H^2}F(\varphi_0) -\frac{2}{3(3\lambda-1)}\frac{\dot\varphi_0}{\chi^4 H^3}(3H\dot\varphi_0+V_{0,\varphi})F(\varphi_0) +\nonumber\\
&-\frac{1}{2(\lambda-1)}\frac{\dot\varphi_0^4}{\chi^4 H^2}-4\frac{\dot\varphi_0^2}{\chi^2} +\frac{V_{0,\varphi\varphi}\dot\varphi_0^2}{\chi^2 H^2} -\frac{1}{\chi^2 H^2}(3H\dot\varphi_0 + V_{0,\varphi})^2+\nonumber\\
&+\frac{5\dot\varphi_0}{\chi^2 H}(3H\dot\varphi_0 + V_{0,\varphi})-
\frac{1}{9(3\lambda-1)^2}\frac{\dot\varphi_0^2}{\chi^6 H^4}F(\varphi_0)^2+3\frac{\dot\varphi_0 V_{0,\varphi}}{H\chi^2}
\end{align}
\end{subequations}

\begin{subequations}
\begin{align}
f_\zeta & = - \frac{1}{3(3\lambda-1)}\frac{\dot\varphi_0^2}{\chi^4 H^3}F(\varphi_0)+\frac{\dot\varphi_0^2}{\chi^2 H} - \frac{\dot\varphi_0}{\chi^2 H^2}(3H\dot\varphi_0+V_{0,\varphi})\\
f_\psi & =  -\frac{1}{3(3\lambda-1)}\frac{\dot\varphi_0^2}{\chi^4 H^3}F(\varphi_0)+\left(4+\frac{3\lambda-1}{\lambda-1}\right)\frac{\dot\varphi_0^2}{\chi^2 H} - \frac{\dot\varphi_0}{\chi^2 H^2}(3H\dot\varphi_0+V_{0,\varphi}) - 6H(3\lambda-1)\\
f_{\zeta\psi} &= f_\zeta\\
\tilde f_{\zeta\psi} &= -\frac{1}{3(3\lambda-1)}\frac{\dot\varphi_0^2}{\chi^4 H^3}F(\varphi_0) + \left(1+\frac{3\lambda-1}{\lambda-1}\right)\frac{\dot\varphi_0^2}{\chi^2 H}-\frac{\dot\varphi_0}{\chi^2 H^2}(3H\dot\varphi_0+V_{0,\varphi})\\
g_{\zeta\psi} & =  \frac{\dot\varphi_0^2}{\chi^2 H^2}
\end{align}
\end{subequations}

\begin{subequations}
\begin{align}
\omega_\zeta &= \frac{V_1(\varphi_0)\dot\varphi_0^2}{\chi^2 H^2}\\
\omega_\psi &= \omega_\zeta - 2\\
\tilde\omega_{\zeta\psi} &= \omega_\zeta\\
\omega_{\zeta\psi} &= \omega_\zeta
\end{align}
\end{subequations}

\begin{subequations}
\begin{align}
d_\psi &= -4\frac{\dot\varphi_0 V_4(\varphi_0)}{\chi^2 H} + 4\frac{V_4(\varphi_0)\dot\varphi_0}{\chi^2 H}-\frac{[V_{4,\varphi}(\varphi_0)+V_2(\varphi_0)]\dot\varphi_0^2}{\chi^2 H^2}-\frac{2}{\chi^2}(8g_2+ 3 g_3)\\
d_{\zeta\psi} &= -4\frac{\dot\varphi_0 V_4(\varphi_0)}{\chi^2 H}+4\frac{V_4(\varphi_0)\dot\varphi_0}{\chi^2 H} -2\frac{[V_{4,\varphi}(\varphi_0)+V_2(\varphi_0)]\dot\varphi_0^2}{\chi^2 H^2}\\
d_{\zeta} &= -\frac{[V_{4,\varphi}(\varphi_0)+V_2(\varphi_0)]\dot\varphi_0^2}{\chi^2 H^2}\\
\tilde d_{\zeta\psi} &= -2\frac{V_6(\varphi_0) \dot\varphi_0^2}{\chi^2 H^2}\\
\tilde d_\psi &= -\frac{V_6(\varphi_0) \dot\varphi_0^2}{\chi^2 H^2}-\frac{2}{\chi^4 }(8g_7-3g_8) \\
\tilde d_\zeta &= -\frac{V_6(\varphi_0) \dot\varphi_0^2}{\chi^2 H^2} 
\end{align}
\end{subequations}

\bibliography{HL}

\end{document}